\documentclass[12pt]{article}
\usepackage{amsmath,amsfonts}
\makeatletter \@addtoreset{equation}{section}

\usepackage{cite}
\usepackage{bm}
\usepackage{dcolumn}
\makeatother
\def\one{{\hbox{ 1\kern-.8mm l}}}
\newcommand{\Dslash}{\not{\hbox{\kern-4pt $D$}}}
\newcommand{\pdslash}{\not{\hbox{\kern-2pt $\partial$}}}
\newcommand{\be}{\begin{equation}}
\newcommand{\bea}{\begin{eqnarray}}
\newcommand{\eea}{\end{eqnarray}}
\newcommand{\ba}{\begin{array}}
\newcommand{\ea}{\end{array}}
\newcommand{\ee}{\end{equation}}
\begin{document}
\begin{titlepage}
\vspace*{1mm}%
\hfill%
\vbox{
    \halign{#\hfil        \cr
           IPM/P-2009/002 \cr
                     } 
      }  
\vspace*{15mm}%
\begin{center}

{{\Large {\bf  Fermions in non-relativistic AdS/CFT correspondence }}}

\vspace*{15mm}
\vspace*{1mm}
{Amin Akhavan$^{a}$, Mohsen Alishahiha$^{b}$, Ali Davody$^{a,b}$ and  
Ali Vahedi$^{a,b}$}

 \vspace*{1cm}
{\it ${}^a$ Department of Physics, Sharif University of Technology \\
P.O. Box 11365-9161, Tehran, Iran}

\vspace*{.4cm}

{\it ${}^b$ School of physics, Institute for Research in Fundamental Sciences (IPM)\\
P.O. Box 19395-5531, Tehran, Iran \\ }

\vspace*{.4cm}
{  amin\_akhavan@mehr.sharif.ir, alishah ,davody, vahedi @ipm.ir} 

\vspace*{2cm}
\end{center}

\begin{abstract}
We extend the non-relativistic AdS/CFT correspondence to the fermionic fields. 
In particular we study the two point function of a fermionic operator in 
 non-relativistic CFTs by making use of a massive fermion propagating in  geometries with Schr\"odinger group 
isometry. Although the boundary of the geometries with Schr\"odinger group isometry differ from that in 
AdS geometries where the dictionary of AdS/CFT is established, using the general procedure of AdS/CFT
correspondence, we see that the resultant two point function
has the expected form for fermionic operators in non-relativistic CFTs, though a non-trivial 
regularization may be needed.

\end{abstract}

\end{titlepage}

\section{Introduction }

In this paper, motivated by general idea of AdS/CFT correspondence \cite{Maldacena:1997re}, we would like to 
study non-relativistic fermions by making use of a gravity description. More precisely we will explore how to compute the two point function of a fermionic operator in a non-relativistic  CFT generalizing the relativistic
one first studied in \cite{Henningson:1998cd}.

Non-relativistic CFT in $d$ dimensions has $d-1$ dimensional Schr\"odinger symmetry, $Sch_{d-1}$. 
The generators of the Schr\"odinger algebra are spatial translations $P_i$, rotations $M_{ij}$, time translation $H$, 
Galilean boosts $K_i$, dilation $D$, number operator $N$ and special conformal transformation $C$. 
The algebra of $Sch_{d-1}$ is given by\footnote{This algebra can be obtained from the relativistic conformal 
algebra in $d+1$ dimensions by a contraction. In other words the 
Schr\"odinger group may be 
thought of as a subgroup of $SO(2,d+1)$ with fixed momentum along the null direction.}
\bea
&&[M_{ij},P_k]=-i(\delta_{ik}P_j-\delta_{jk}P_i),\;\;\;\;\;\;\;\;\;\;\;\;\;\;\;\;\;
[M_{ij},K_k]=-i(\delta_{ik}K_j-\delta_{jk}K_i),\cr
&&[M_{ij},M_{kl}]=-i\delta_{ik}M_{jl}+{\rm perms},\;\;\;\;\;\;\;\;\;\;\;\;\;\;[P_i,K_j]=-iN\delta_{ij}\cr
&&[D,P_i]=-iP_i,\;\;\;\;\;\;\;\;[D,K_i]=iK_i,\;\;\;\;\;\;\;\;\;[D,H]=-2iH,\cr
&&[C,P_i]=iK_i,\;\;\;\;\;\;\;\;\;\;\;[C,D]=-2iC,\;\;\;\;\;\;\;[C,H]=-iD.
\eea
Since $[D,N]=0$ one may diagonalize them simultaneously leading to the fact that representations of 
the Schr\"odinger algebra may be labeled by two numbers; dimension $\Delta$ and a number $M$ which are
the eigenvalues of $D$ and $N$, respectively.

Following the relativistic AdS/CFT correspondence one expects that  if there are gravity duals to  
non-relativistic CFTs, the isometry of the relevant geometry must be Schr\"odinger group. In fact
such gravity duals exist and the corresponding 
geometry is given by 
\be\label{sonic}
ds^2=\rho^2(-\rho^2{dt^2}-2dtd\xi+dx_i^2)+\frac{d\rho^2}{\rho^2},\;\;\;\;\;i=1,\cdots,d-1,
\ee
which could be thought of as a solution of a $d+2$ dimensional gravity coupled to massive
gauge field \cite{Son:2008ye,Balasubramanian:2008dm}. 

This geometry has been used to study different features of non-relativistic CFT. 
In particular utilizing a propagating massive scaler field in the 
bulk geometry \eqref{sonic}  the corresponding point function in the 
dual non-relativistic CFT has been calculated \cite{Balasubramanian:2008dm}. 
Although the geometry \eqref{sonic} has a one dimensional
time-like boundary\footnote{$g_{tt}$ grows faster than the other metric components
for large $\rho$.}, it has been used to compute two point functions of $d$ dimensional non-relativistic CFT.
In other words, in order to compute the correlation function in the non-relativistic AdS/CFT it was 
implicitly assumed in \cite{Balasubramanian:2008dm} that the procedure is the same as that in 
the relativistic case \cite{Gubser,Witten} where the gravity description is given in terms of AdS geometry 
which has a well defined boundary\footnote{Such an assumption has generically been accepted in the 
literature. See for example\cite{{Chen:2008ad},{Kachru:2008yh},{Kovtun:2008qy},{Lin:2008pi},{Akhavan:2008ep},
{Taylor:2008tg},{Danielsson:2008gi},{Pal:2009yp},{Alishahiha:2009hg}}}. 
Indeed following this assumption we will get the expected two point function of scalars in the
non-relativistic CFT, though a non-trivial regularization seems to be needed \cite{Balasubramanian:2008dm}. 
One then may wonder if the need for such a regularization is due to the peculiar boundary of the geometry
\eqref{sonic}. 

The aim of this paper is to further explore the AdS/CFT correspondence in the context of non-relativistic CFTs.
We study correlation functions of fermionic operators in the non-relativistic CFT
by making use of fermions propagating in the bulk geometry given by \eqref{sonic}.
More precisely we will consider a propagating massive fermion in the geometry \eqref{sonic}
and compute the value of the action with a proper boundary term for a classical solution of the equation of motion.
Then we shall identify this value with the generating function of a fermionic operator in the non-relativistic
CFT. 

Since the non-relativistic AdS/CFT correspondence relates two theories with dimensions $d$ and $d+2$, an immediate difficulty we face when we are considering fermions in this context is that the number of degrees of freedom of the
spinors do not match for these two theories. Therefore there must be a condition which projects out half 
of the degrees of freedom. Actually this is the case. In fact the correspondence is smart enough to
automatically project out half of the degrees of freedom due to its particular {\it boundary interaction}.
We note, however, that although following the general procedure of AdS/CFT correspondence we will 
get the expected fermionic correlation function, a non-trivial
regularization is needed to make the result finite. 

The paper is organized as follows. In the next section we will study fermions in $d$ dimensional
non-relativistic CFT by making use of free relativistic fermions in $d+1$ dimensions.
In particular using the field theory method we compute two point function of the non-relativistic fermions. 
In section three we consider gravity description of the non-relativistic
fermions where the fermion's two point function is found via gravity calculations. 
The last section is devoted to discussions.

\section{Non-relativistic fermions}

In this section we review non-relativistic fermions in $d$ dimensions and compute their two 
point function\footnote{ The non-relativistic fermion  has been studied in the
context of supersymmetric Schr\"odinger symmetry in \cite{Henkel:2005dj}.}.
First of all we note that  the $d$ dimensional Galilean conformal symmetry may be obtained from 
conformal symmetry in $d+1$ dimensions \cite{{Burdet:1977qw},{Duval:1984cj},{D1},{D2},{Henkel:2003pu},
{Duval:1993hs},{Sakaguchi:2008rx}
,{D3}}.
Therefore to proceed we will start from the action of a massless relativistic fermion in $d+1$ dimensions whose 
symmetry is $SO(2,d+1)$, the conformal group in $d+1$ dimensions, then compactify it to $d$ dimensions along a 
null direction. In this way we will end up with the action of a non-relativistic fermion.

To be specific we will consider $d=4$, though it can be easily generalized to other dimensions.
The action of $d+1$ dimensional massless fermion is\footnote{Here we work in $(-,+,+,+,+)$ signature.}
\begin{equation}\label{act}
 S= \int d^{5}x\; \bar{\psi}i\gamma^{\mu}\partial_{\mu}\psi.
\end{equation}
Let us decompose the coordinates as $(t,\xi,x_i),\;i=1,2,3$, where the light like coordinates $(t,\xi)$
are defined by
\begin{eqnarray}
t=\frac{1}{\sqrt{2}}(x^{0}+x^{4}),\;\;\;\;\;\;\;\;\;\;\xi=\frac{1}{\sqrt{2}}(x^{0}-x^{4}).
\end{eqnarray}
Accordingly the gamma matrices can also be decomposed along the light like and spatial directions
\begin{eqnarray}
\gamma_{t}=\frac{1}{\sqrt{2}}(\gamma_{0}+\gamma_{4}),\;\;\;\;\;\;
\gamma_{\xi}=\frac{1}{\sqrt{2}}(\gamma_{0}-\gamma_{4}),\;\;\;\;\;\;\gamma_i\;\;{\rm for}\;\;i=1,2,3.
\end{eqnarray}
Using this notation the action (\ref{act}) reads
\begin{eqnarray}
 S=\frac{i}{\sqrt{2}}
 \int d^{3}x d\xi dt\;  \psi^{\dag}\bigg{(}\gamma_{\xi}\gamma_{t}\partial_{\xi}+\gamma_{t}\gamma_{\xi}\partial_{t}
 -(\gamma_{\xi}+\gamma_{t})\gamma_{i}\partial_{i}\bigg{)}\psi.
\end{eqnarray}
Consider a single mode with definite momentum in the null direction $\xi$. So that
$\psi(t,\xi,x)=e^{iM\xi}\psi_{M}(t,x)$. Therefore the action becomes
\begin{eqnarray}
S= \frac{i}{\sqrt{2}}
 \int d^{3}x dt\;
\psi_{-M}^{\dag}\bigg{(}iM\gamma_{\xi}\gamma_{t}+\gamma_{t}\gamma_{\xi}\partial_{t}
 -(\gamma_{\xi}+\gamma_{t})\gamma_{i}\partial_{i}\bigg{)}\psi_{M}.
 \end{eqnarray}
Using the explicit representation for gamma matrices and setting
\be
\psi_M(x,t)=e^{i(Et+k_ix_i)} \left( \begin{array}{c}  \phi \\ \chi \\ \end{array}\right),
\ee
the equation of motion obtained from the above action is given by
\begin{equation}
\left(
\begin{array}{cc}
  2E & -i\sqrt{2} {\sigma}_i {k}_i \\
  i\sqrt{2} {\sigma}_i{k}_i & 2M \\
\end{array}
\right)
\left(\begin{array}{c} \phi \\   \chi \\ \end{array} \right)=0,
\end{equation}
where $\phi$ and $\chi$ are two component Weyl spinors and $\sigma_i$'s are Pauli's matrices.
This is the equation of motion for non-relativistic fermion which was studied in \cite{Hagen:1972pd} where
it was shown that the equation is invariant under the non-relativistic conformal symmetry. Indeed under
the scaling  $t\rightarrow \lambda^2 t,\;x_i\rightarrow \lambda x_i$,
the $\phi$ and $\chi$ have scaling dimension $-\frac{3}{2}$ and $-\frac{5}{2}$, respectively.

The most general solution of the above equation is
\begin{eqnarray}
\psi_{M}(t,x)= \int \frac{d^{3}k}{(2\pi)^{3/2}}\; e^{i(\frac{k^{2}}{2M}t+k_i {x}_i)}\left(
\begin{array}{c}
  \sum_{s=1}^2a^{s}(k)\;\eta^{s} \\ \\
  -i\frac{\sigma_i{k}_i}{\sqrt{2}M}\; \sum_{s=1}^2a^{s}(k)\;\eta^{s}
\end{array}
\right),
\label{sol}
\end{eqnarray}
where
\be
\eta^{1}=\left(\begin{array}{c} 1 \\ 0 \\ \end{array}\right),\;\;\;\;
\eta^{2}=\left( \begin{array}{c}   0 \\   1 \\ \end{array} \right),\;\;\;\;
\{a^{s}_M(k),
{a^\dagger}^{r}_{-M}(k')\}=(2\pi)^{3/2}\delta^{sr}\delta^3(k-k^{'}).
\label{anti}
\ee

Now we are ready to find two point function of the non-relativistic fermions. Using the
explicit solution (\ref{sol}) one finds
\begin{equation}
    \langle\; \psi_M(t,x)\bar\psi_{-M}(0,0)\;\rangle=\int \frac{d^{3}k}{(2\pi)^{3/2}}\;
    \frac{d^{3}k^{'}}{(2\pi)^{3/2}}\; e^{i(\frac{k^{2}}{2M}t+k_i {x}_i)}\; {\cal M},
\end{equation}
where
\be
{\cal M}=\left(
\begin{array}{cc}
 \frac{{\sigma }_i{k'}_i}{\sqrt{2}M}\; \langle 0|\; a^{s}(k)a^{r\dag}(k^{'})\;|0\rangle\;
 \eta^{s}\eta^{r\dag}&\;\;\;\;
 -i\;\langle 0|\;a^{s}(k)a^{r\dag}(k^{'})\;|0\rangle\; \eta^{s}\eta^{r\dag}
    \\ &\\ -i\frac{\sigma_i\sigma_jk_ik'_j}{2M}\;\langle 0|\;a^{s}(k)a^{r\dag}(k^{'})\;|0\rangle\;\eta^{s}\eta^{r\dag}
  &\;\;\;\; -\frac{\sigma_ik_i}{\sqrt{2}M}\;\langle 0|\;a^{s}(k)a^{r\dag}(k')\;|0\rangle\;\eta^{s}\eta^{r\dag} \\
\end{array}
\right)
\ee
It is easy to evaluate the expectation values of $a(k)$'s using the anticommutator in (\ref{anti}).
Then performing the integration over $k'$ one arrives at
\begin{equation}
 \langle\; \psi_M(t,x)\bar\psi_{-M}(0,0)\;\rangle=-\int \frac{d^{3}k}{(2\pi)^{3/2}}
  \left(
\begin{array}{cc}
  -\frac{\sigma_ik_i}{\sqrt{2}M}&  i \\ &\\
  i\frac{k^{2}}{2M^2}&\frac{\sigma_ik_i}{\sqrt{2}M}
\end{array}
\right)\;e^{i(\frac{k^{2}}{2M}t+k_ix_i)}.
\label{1}
\end{equation}
which can be recast to the following form
\be\
\langle\; \psi_M(t,x)\bar\psi_{-M}(0,0)\;\rangle=\frac{i}{\sqrt{2}M}
\bigg{(}iM\gamma_{t}+\gamma_{\xi}\partial_{t}-\gamma_{i}\partial_{i}\bigg{)} G(t,x;0,0),
\label{ee}
\ee
where
\begin{equation}
G(t,x;0,0)= \int \frac{d^{3}k}{(2\pi)^{3/2}} \;e^{i(\frac{k^{2}}{2M}t+k_ix_i)}=e^{\frac{3\pi i}{4}} 
\left(\frac{M}{t}\right)^{3/2}e^{-\frac{iM x^2}{2t}}
\end{equation}
is the Green function of the non-relativistic scalar field propagating in four
dimensions \cite{{Balasubramanian:2008dm},{Goldberger:2008vg},{Henkel:2003pu},{Henkel:2005dj}}.

\section{Gravity description}

In this section we will study non-relativistic fermions in $d$ dimensions by making use of a 
$d+2$ dimensional gravity with Schr\"odinger group isometry. 
The corresponding gravity solution is given in \eqref{sonic}. Setting $r=\frac{1}{\rho}$ the metric
\eqref{sonic} reads
\be
ds^2=-\mu^2 \frac{dt^2}{r^{4}}+\frac{2dtd\xi+dx_i^2+dr^2}{r^2}.
\label{back} 
\ee
Here we added the parameter $\mu$ which parametrizes the deviation from 
$AdS_{d+2}$ geometry\footnote{We note, however, that due to the scaling symmetry $t\rightarrow t/\beta,\;
\xi\rightarrow \beta \xi$ of the geometry for $\mu\neq 0$ it can be rescaled to any value. On the other hand since 
$\partial_\xi$ may be identify with number operator of the dual non-relativistic CFT, $\xi$ direction 
should be periodic; $\xi\equiv\xi+2\pi L$. It is then natural to define the dimensionless parameter 
$\mu L$ which may parametrize the non-relativistic nature of the theory\cite{Adams:2008wt}.
Nevertheless we prefer to set it as a free parameter to follow the effects of deformation.
For a nice discussion concerning this point see \cite{Herzog:2008wg}. }.

The aim is to solve the Dirac equation in this background with an appropriate boundary condition.
Following the  AdS/CFT correspondence \cite{{Gubser},{Witten}} this may be used to evaluate two point function of 
an operator which would be dual to the bulk fermion. Due to the symmetry of the proposed background 
(\ref{back}) one expects that the obtained two point function corresponds to the two point 
function of a fermionic operator in the dual non-relativistic CFT.

The Dirac equation in the background (\ref{back}) is given by
\begin{eqnarray}\label{dirac}
\left(r\Gamma_{\hat{t}} \partial_{\xi}+r\Gamma_{\hat{\xi}} \partial_{t}+
r\Gamma_{\hat{i}}
\partial_{i}+r\Gamma_{\hat{r}} \partial_{r}+\frac{\mu^2}{2r}\Gamma_{\hat{\xi}} \partial_{\xi}-
 \frac{d+1}{2} \Gamma_{\hat{r}}-m\right) \Psi(x_i,t,\xi,r)
 =0, 
\end{eqnarray}
where $\Gamma_{\hat A}$'s are $(d+2)$-dimensional gamma matrices obeying\footnote{The hatted 
indices denote the coordinates of the tangent space.}
\begin{eqnarray}
 \{\Gamma_{\hat \xi},\Gamma_{\hat t}\}=2,\;\;\{\Gamma_{\hat \xi},\Gamma_{\hat \xi}\}=0,\;\;
 \{\Gamma_{\hat t},\Gamma_{\hat t}\}=0,\;\;\{\Gamma_{\hat i},\Gamma_{\hat j}\}=2
\eta_{\hat i,\hat j},\;\;\{\Gamma_{\hat i},\Gamma_{\hat t,\hat \xi}\}=0.
\end{eqnarray}

We note, however, that it is not straightforward to solve the the Dirac equation (\ref{dirac}) exactly. 
Nevertheless since we are interested in a solution near $r=0$, one may try to find the solution 
perturbatively near $r=0$. 
To proceed we multiply the Dirac 
equation \eqref{dirac} by 
$$r\Gamma_{\hat{t}}
\partial_{\xi}+r\Gamma_{\hat{\xi}} \partial_{t}+ r\Gamma_{\hat{i}}
\partial_{i}+r\Gamma_{\hat{r}} \partial_{r}+\frac{\mu^2}{2r}\Gamma_{\hat{\xi}}
\partial_{\xi}$$
leading to the following Laplace equation 
\begin{equation}\label{sd}
\left(\textbf{D}^2+m\Gamma_{\hat{r}}-\frac{\mu^2}{r}\Gamma_{\hat{r}}\Gamma_{\hat{\xi}}\partial_{\xi}
\right) \Psi(x_i,t,\xi,r)=0,
\end{equation}
where
\begin{eqnarray}
\textbf{D}^2=\
r^2\partial_{r}^{2}-r(d+1)\partial_{r}+r^2(2\partial_{\xi}\partial_{t}+\partial_i^2)
+\left(\frac{(d+1)^2}{4}+\frac{d+1}{2}-m^2+\mu^2
\partial_{\xi}^{2}\right),
\end{eqnarray}
To find the solution  it is useful to Fourier transform 
over $t$ and $x_i$ directions. Denoting by $M$ the momentum along the null direction $\xi$ one has
\begin{eqnarray}
\Psi(t,x_i,\xi,r)=e^{iM\xi}\int \frac {d\omega d^{d-1}q}{(2\pi)^{d/2}}
\;\;e^{i\omega t+i q.x}\;\Psi_M(q,\omega,r),
\end{eqnarray}
by which the equation (\ref{sd}) reads 
\begin{equation}\label{d2}
(\textbf{D}^2+m\Gamma_{\hat r}-\frac{i\mu^2 M}{r}\Gamma_{\hat r}\Gamma_{\hat \xi})\Psi_M(q,\omega, r)=0,
\end{equation}
where
\begin{eqnarray}
\textbf{D}^2=\ r^2\partial_{r}^{2}-r(d+1)\partial_{r}-r^2 k^{2}
+(\frac{(d+1)^2}{4}+\frac{(d+1)}{2}-m^2-\mu^2 M^{2}),
\end{eqnarray}
with $k^2=2M\omega+q^2$.  On the other hand since $\Gamma_{\hat \xi}^2=0$ one finds
\begin{equation}\label{sd1}
(\textbf{D}^2-m\Gamma_{r})\Gamma_{\hat{\xi}}\Psi_M(k,r)=0.
\end{equation}
This equation can be exactly solved to find $\Gamma_{\hat{\xi}}\Psi_M(k,r)$. To do so, taking into account
that $\Gamma_{\hat r}^2=1$, one can decompose $\Psi_M(k,r)$ in terms of eigenvectors of $\Gamma_{\hat{r}}$ as follows
\be
\Psi_M(k,r)=\Psi^{+}_M(k,r)+\Psi^{-}_M(k,r),\;\;\;\;\;\;\;\;\;\;\;\;\;\Gamma_{\hat{r}}\Psi^\pm_M=\pm\Psi^\pm_M\ .
\ee
So that $\Gamma_{\hat{r}}\Gamma_{\hat{\xi}}\Psi^{\pm}_M(k,r)=\mp\Gamma_{\hat{\xi}}\Psi^{\pm}_M(k,r)$.
By making use of the above decomposition the 
equation \eqref{sd1} reduces to two equations for $\Gamma_{\hat{\xi}}\Psi_M(k,r)^{\pm}$
\begin{equation}\label{d3}
(\textbf{D}^2+m)\Gamma_{\hat{\xi}}\Psi^{+}_M(k,r)=0,\;\;\;\;\;(\textbf{D}^2-m)\Gamma_{\hat{\xi}}\Psi^{-}_M(k,r)=0,
\end{equation}
whose solutions are 
\begin{equation}\label{11}
\Gamma_{\hat{\xi}}\Psi^{+}_M(k,r)=r^{\frac{d}{2}+1}\mathbf{K}_{\nu^{-}}(kr)
\Gamma_{\hat{\xi}}{\bf u}_M(k),\;\;\;\;\;\Gamma_{\hat{\xi}}\Psi^{-}_M(k,r)=r^{\frac{d}{2}+1}\mathbf{K}_{\nu^{+}}(kr)
\Gamma_{\hat{\xi}}{\bf v}_M (k).
\end{equation}
Here ${\bf v}_M (k)$ and ${\bf u}_M (k)$ are constant spinors, $\mathbf{K}_{\nu}(z)$ is the Bessel function and 
\begin{equation}
    \nu^{\pm}=\sqrt{(m\pm\frac{1}{2})^{2}+\mu^2 M^2}.
\end{equation}
Plugging the solutions (\ref{11}) into the equations (\ref{d2}) we get  
\bea
(\textbf{D}^2+m)\Psi^{+}_M(k,r)&=&\frac{i\mu^2
M}{r}r^{\frac{d}{2}+1}\mathbf{K}_{\nu^{+}}(kr)
\Gamma_{\hat{\xi}}{\bf v}_M(k),\cr && \cr\;\;\;\;\;(\textbf{D}^2-m)\Psi^{-}_M(k,r)&=&\frac{-i\mu^2
M}{r}r^{\frac{d}{2}+1}\mathbf{K}_{\nu^{-}}(kr)
\Gamma_{\hat{\xi}}{\bf u}_M(k).
\eea
The most general solution to the above equations are
\bea\label{12}
\Psi^{+}_M(k,r)&=&r^{\frac{d}{2}+1}\mathbf{K}_{\nu^{-}}(kr){\bf u}_M(k)+f(k,r)\;\Gamma_{\hat{\xi}}{\bf v}_M(k)
\cr &&\cr
\Psi^{-}_M(k,r)&=&r^{\frac{d}{2}+1}\mathbf{K}_{\nu^{+}}(kr){\bf v}_M(k)+g(k,r)\;\Gamma_{\hat{\xi}}{\bf u}_M(k)
\eea
where $f(k,r)$ and $g(k,r)$ are solutions of the following differential
equations
\begin{equation}
(\textbf{D}^2+m)f(k,r)=i\mu^2 M
 r^{\frac{d}{2}}\mathbf{K}_{\nu^{+}}(kr)
,\;\;\;\;\;(\textbf{D}^2-m)g(k,r)=-i\mu^2 M
r^{\frac{d}{2}}\mathbf{K}_{\nu^{-}}(kr).
\end{equation}
In general it is not possible to solve the above equations exactly getting a closed form for 
the solution. Actually we do not even need to solve the equations exactly. The only thing we need
is the asymptotic  behavior of the solutions near $r=0$. Therefore we will solve the above equations order 
by order near $r=0$. Doing so, in the lowest order we arrive at
\bea
f(k,r)&=&a k^{-\nu^+}
r^{\frac{d}{2}-\nu^{+}}+a'r^{\frac{d}{2}-\nu^{+}+2}+\cdots\cr
g(k,r)&=&b k^{-\nu^-}
r^{\frac{d}{2}-\nu^{-}}+b'r^{\frac{d}{2}-\nu^{+}+2}+\cdots
\eea
where $a,b,a',b',\cdots$ are numerical factors. In fact for our purpose we only need the first term
in the expressions of $g$ and $f$ where the corresponding coefficients are given by
\begin{equation}\label{14}
a =i\mu^2\frac{2^{\nu^{+}-2}
M\Gamma(\nu^{+})}{\nu^{+}+m+\frac{1}{2}}\;\;\;,\;\;\;\;b
=-i\mu^2\frac{2^{\nu^{-}-2}
M\Gamma(\nu^{-})}{\nu^{-}-m+\frac{1}{2}}.
\end{equation}

Note that so far we have been trying to solve the equation \eqref{d2} which is actually the square of the Dirac 
equation \eqref{dirac}. Therefore a solution of a equation \eqref{d2} is not necessarily a solution of the
Dirac equation \eqref{dirac} as well. 
To find a solution of the Dirac equation one needs to plug
the solution of \eqref{d2} into the Dirac equation which in general leads to an extra condition on the
constant spinors. In particular in our case we find
\begin{equation}\label{cons}
 {\bf v}_M=\frac{-i}{2M}\;\; (i\kappa_{\hat a}\Gamma_{\hat a})\Gamma_{\hat \xi}{\bf v}_M,\;\;\;\;\;\;\;{\bf u}_M=\frac{-i}{2M}\;\; (i\kappa_{\hat a}\Gamma_{\hat a})\Gamma_{\hat \xi}{\bf u}_M,
\end{equation}
where $(\kappa_{\hat t},\kappa_{\hat \xi},\kappa_{\hat i})=(M,\omega,q_i)$. 
It is worth mentioning that in these expressions there is a term, $\omega\Gamma_{\hat \xi}$, which
has naively zero contribution due to the fact that $\Gamma_{\hat \xi}^2=0$.
We note, however, that as we we will see the physical spinors which inter in the dictionary of 
AdS/CFT correspondence in the non-relativistic case are $(\Gamma_{\hat \xi}{\bf u}_M,\Gamma_{\hat \xi}{\bf v}_M)$ 
and not $({\bf u}_M,{\bf v}_M)$ themselves. Therefore it is important to keep this term too.
 
Similarly going through the above procedure for $\overline{\Psi}_{-M}(k,r)$ one finds
\bea\label{13}
 \overline{\Psi}^{+}_{-M}(k,r)&=&r^{\frac{d}{2}+1}\mathbf{K}_{\nu^{+}}(kr)
\overline{\bf u}_{-M}(k)+g(k,r)\overline{\bf v}_{-M}(k)\Gamma_{\hat{\xi}}\cr &&\cr
\overline{\Psi}^{-}_{-M}(k,r)&=&r^{\frac{d}{2}+1}\mathbf{K}_{\nu^{-}}(kr)
\overline{\bf v}_{-M}(k)+f(k,r)\overline{\bf u}_{-M}(k)\Gamma_{\hat{\xi}}.
\eea

Now we have all ingredients to establish the AdS/CFT correspondence for the non-relativistic fermions.
First of all we note that since in the non-relativistic case it is believed that 
$d$ dimensional non-relativistic CFT is
described by a $d+2$ dimensional gravity, a Dirac spinor in the bulk is $2^{[\frac{d}{2}+1]}$
dimensional, though in the dual non-relativistic field theory it has $2^{[\frac{d}{2}]}$ dimensions.
Therefore a priori it is not clear how to match the degrees of freedom in the bulk with that in the 
dual theory. So if the AdS/CFT works for the non-relativistic case, one would 
expect that there should be a constraint on the spinor in the bulk reducing its degrees of freedom with 
a factor of two. In fact, as we have already anticipated, the correspondence is clever enough 
to project out half of the degrees of freedom. To see this we note that at $r \rightarrow 0$ 
we have 
\bea
\lim_{r\rightarrow 0} \Psi_M(k,r)\sim
r^{\frac{d}{2}-\nu^{+}}\Gamma_{\hat{\xi}}{\bf v}_M,\;\;\;\;\;
\lim_{r\rightarrow0}\overline{\Psi}_{-M}(k,r)\sim
r^{\frac{d}{2}-\nu^{+}}\overline{\bf u}_{-M} \Gamma_{\hat{\xi}},
\eea
showing that if we want to follow the general procedure of AdS/CFT correspondence
 interpreting the asymptotic behavior of 
the spinor at $r\rightarrow 0$ as the source for the dual operator in the non-relativistic field theory,
the source should be given by $\Gamma_{\hat{\xi}}{\bf v}_M$ and $\overline{\bf u}_{-M} \Gamma_{\hat{\xi}}$
rather than ${\bf v}_M$ and $\overline{\bf u}_{-M}$. This is exactly what we need.

Indeed the spinors $\Gamma_{\hat{\xi}}{\bf v}_M$ and 
$\overline{\bf u}_{-M} \Gamma_{\hat{\xi}}$ have the appropriate independent
degrees of freedom to be identified as the sources for the dual operator $\psi$ in non-relativistic CFT.
More precisely to proceed we assume the following coupling in the non-relativistic CFT 
\begin{equation}\label{25}
\mathrm{Z}_{CFT}=\bigg{\langle}\;\exp\left[\int
d^dx\left(\overline{\psi}_{-M}\Gamma_{\xi}{\bf v}_M
+\overline{\bf u}_{-M}\Gamma_\xi \psi_M\right)\right]\bigg{\rangle}.
\end{equation}
Here $\psi_M$ is a $2^{[d/2]}$ dimensional spinor in the non-relativistic CFT appropriately embedded into
$2^{[d/2+1]}$ dimensional spinor representation by doubling the $d$ dimensional spinor.

Therefore to evaluate the two point function $\langle \psi_M\overline{\psi}_{-M}\rangle$ one needs
to perform the variations with respect to $\overline{\bf u}_{-M}\Gamma_{\hat \xi}$ and $\Gamma_{\hat \xi}{\bf v}_M$.
On the other hand we would like to identify the partition partition function $\mathrm{Z}_{CFT}$ of the 
non-relativistic CFT with 
${\exp}(I_{AdS})$, where $I_{AdS}$ is the boundary term in the bulk action given by
\be\label{boundaryterm}
I_{AdS}=\int d\xi dtd^{d-1}x\sqrt{g}\;\overline{\Psi}(t,x,r,\xi)\Psi(t,x,r,\xi).
\ee
Going to the momentum space and for $r=\epsilon$ with $\epsilon$ being an infinitesimal
number, one gets  
\begin{equation}
I_{AdS}=
\epsilon^{-d-1}\int\frac{
 d\omega d^{d-1}q}{(2\pi)^{d/2}}\bigg{[}\overline{\Psi}^{+}_{-M}(k,\epsilon)\Psi^{+}_{M}(k,\epsilon)
 +\overline{\Psi}^{-}_{-M}(k,\epsilon)\Psi^{-}_{M}(k,\epsilon)\bigg{]}.
\end{equation}
On the other hand using the solutions \eqref{12} and \eqref{13} the integrand in leading order reads 
\be
\bigg{[}\overline{\Psi}^{+}_{-M}(k,\epsilon)\Psi^{+}_{M}(k,\epsilon)
 +\overline{\Psi}^{-}_{-M}(k,\epsilon)\Psi^{-}_{M}(k,\epsilon)\bigg{]} 
\approx i\mu^2 C\epsilon^{d+1-2\nu^{+}}k^{-2\nu^{+}}\overline{\bf u}_{-M}(k)\Gamma_{\hat \xi}{\bf v}_{M}(k),
\ee
where 
$
C=\frac{2^{2\nu^+-2} M \Gamma^2(\nu^+)}{\nu^++m+1/2}$. Therefore utilizing the equation \eqref{cons} we arrive at 
\bea
I_{AdS}&=& \frac{\mu^2C}{2M} 
\epsilon^{-2\nu^+}\int\frac{dtd^{d-1}x}{(2\pi)^{d/2}} \;\frac{dt' d^{d-1}x'}{(2\pi)^{d/2}}
\bigg{[}\cr &&\int \frac{d\omega d^{d-1}q}
{(2\pi)^{d/2}}\;\;\;e^{iq.(x-x')+i\omega(t-t')}k^{-2\nu^{+}}
\overline{\bf u}_{-M}(x',t')
   \Gamma_{\hat \xi}(i\kappa_{\hat a}\Gamma_{\hat a})\Gamma_{\hat \xi}{\bf v}_{M}(x,t)\bigg{]}
\eea
which can be used to find the two point function as follows
\be
\langle \psi_M(x,t)\overline{\psi}_{-M}(0,0)\rangle = \frac{\mu^2 C}{2M} \epsilon^{-2\nu^+}\int \frac{d\omega d^{d-1}q}
{(2\pi)^{d/2}}\;\;\;e^{iq.x+i\omega t}k^{-2\nu^{+}}
\; (i\kappa_{\hat a}\Gamma_{\hat a}).
\ee
Performing the integration one finds
\be\label{two}
\langle\psi_M(x,t)\overline{\psi}_{-M}(0,0)\rangle= 
\frac{i\mu^2 B e^{\frac{3\pi i}{4}}}{\sqrt{2}M}\;\epsilon^{-2\nu^+}\bigg(iM\Gamma_{\hat t}+\Gamma_{\hat \xi}\partial_{t}
+\Gamma_{\hat i}\partial_i\bigg)\;\bigg(t^{-\Delta}e^{\frac{iMx^{2}}{2t}}\bigg)
\ee
where $\Delta=\frac{d+1}{2}-\nu^+$. Here $B$ is a numerical factor which can be absorbed in the definition of the 
coupling constant of the boundary term \eqref{boundaryterm}.

As we see this expression has the right form to be identified with the two point function of the 
fermion in the non-relativistic CFT. 
Indeed it is compatible with  \eqref{ee} for $\nu^+=1\; ( m=-1/2\pm\sqrt{1-\mu^2M^2})$\footnote
{Note that due to the definition of $\psi_M$ the resultant two point function \eqref{two} is
two copies of \eqref{ee}. Note also that these two equations are written in different signatures.}. 
We note, however, that it is naively divergent as $\epsilon\rightarrow 0$. This is exactly the same 
singular behavior appeared in the bosonic case \cite{Balasubramanian:2008dm}. Although we do not have
a clear understanding for this behavior, one may wonder that this singularity may be
removed by regularizing the fermionic operators in the non-relativistic CFT.

\section{Discussions}

In this paper we have studied fermions in non-relativistic CFT by making use of a gravity 
description\footnote{ For a study of fermions in the context of dS/CFT correspondence see \cite{Loran:2004fu}.}.
The 
gravity description is given by a geometry with Schr\"odinger isometry. We note, however, that 
although $d+2$ dimensional geometries with Schr\"odinger isometry have one dimensional time-like boundary, 
it may be used to study non-relativistic CFT in $d$ dimensions. Therefore unlike the relativistic
case  where the CFT lives on the boundary of AdS geometry, a priori, it is not
obvious where the non-relativistic CFT is living. As a result it is not clear how to 
extend the dictionary of AdS/CFT correspondence to non-relativistic CFT. 

Nevertheless using the general rules of the relativistic AdS/CFT correspondence we have been able
to obtain two point function for non-relativistic fermions which agrees with that in 
non-relativistic CFT. Of course the resultant two point function is not finite and a non-trivial
regularization seems to be needed. This might be related to the fact that the interested geometry 
has one dimensional time like boundary though it has been used to study $d$ dimensional non-relativistic
CFT. It would be interesting to understand this point better. 

An alternative way to find a gravity dual for non-relativistic CFT has been considered in 
\cite{{Goldberger:2008vg},{Barbon:2008bg}}. 
It was proposed that in order to study non-relativistic CFT in the context of gauge-gravity 
duality one may start from an AdS bulk geometry and break the conformal symmetry of the boundary theory to non-relativistic conformal symmetry by imposing a specific boundary condition for the bulk fields. 
Using this approach the two point function of scalar fields has been computed which is is 
the same form as that in \cite{Balasubramanian:2008dm} up to a shift in the mass of the bulk field.
For further discussions concerning the comparison between these two proposals see \cite{Herzog:2008wg}.

It is worth to apply this method to the fermionic fields and compare the result with that in 
previous section. Since in this case we work with AdS geometry one may borrow the results
which already exist in the literature. More precisely following   
\cite{{Henningson:1998cd},{Mueck:1998iz}} we have to solve the Dirac equation in the $AdS_{d+1}$ geometry
in the light-cone coordinates
\begin{eqnarray}
\bigg(r(\gamma_{\hat t}\partial_{ \xi} +\gamma_{\hat \xi}\partial_{t}
+\gamma_{\hat r}\partial_{r}+\gamma_{\hat i}\partial_{i})-\frac{d+1}{2}\gamma_{\hat r}-m\bigg)\Psi=0.
\label{e30}
\end{eqnarray}
The boundary condition we need to impose is 
\begin{equation}\label{rrr}
    \Psi(t,x_i,\xi,r)= e^{-iM\xi}\Psi_{M}(t,x_i,r),
 \end{equation}
which breaks the conformal symmetry of the boundary theory to non-relativistic conformal symmetry.
Following the notation of \cite{Mueck:1998iz} we note that the two point function has the same expression as that 
in relativistic case except the momentum along $\xi$ direction is fixed. Therefore 
we get
\be
\langle\psi_M(x,t)\overline{\psi}_{-M}(0,0)\rangle=\lim_{\epsilon\rightarrow 0}
i\epsilon^{-2m}\int \frac{d\omega d^{d-1}q}{(2\pi)^{d}}\;
\frac{\kappa_{\hat a}\gamma_{\hat a}}{k} \;e^{i(\omega t+q\cdot x)}\;
\frac{K_{m-\frac{1}{2}}(k\epsilon)}{K_{m+\frac{1}{2}}(k\epsilon)}
\ee
leading to
\be
\langle\psi_M(x,t)\overline{\psi}_{-M}(0,0)\rangle=C  
(i\gamma_{\hat t}M+\gamma_{\hat \xi}\partial_{t}+\gamma_{\hat i}\partial_{i})
\left(\frac{2M}{t}\right)^{\Delta}e^{\frac{iMx^2}{2t}}.
\ee
where $\Delta=m+d/2$. This has the same form as that in previous section 
except the dimension of the operators is different. Of course 
having the same expression is not surprising as the two point function can be 
fixed by the symmetry. The non-trivial point is that these two approaches lead to
different scaling dimensions for operators. It would be extremely interesting to 
explore any possible connection between these two approaches. 

\section*{Acknowledgments}
We would like to thank F. Ardalan, R. Fareghbal, A. E. Mosaffa and  S. Rouhani for discussions on the 
different aspects of non-relativistic AdS/CFT correspondence. This work is supported in 
part by Iranian TWAS chapter at ISMO.



\end{document}